\documentclass{article}

\usepackage{cite} 
\usepackage{geometry} 

\title{Stateful Hash-Based Signature (SHBS) Benchmark Data for XMSS and LMS}
\author{
  Brian Romansky \and
  Thomas Mazzuchi \and
  Shahram Sarkani
}
\date{\today}

\begin{document}

\maketitle

\begin{abstract}
  The National Institute of Standards and Technology (NIST) has recommended the use of stateful hash-based digital signatures for long-term applications that may require protection from future threats that use quantum computers.  XMSS and LMS, the two approved algorithms, have multiple parameter options that impact digital signature size, public key size, the number of signatures that can be produced over the life of a keypair, and the computational effort to validate signatures.  This collection of benchmark data is intended to support system designers in understanding the differences among the configuration options.
\end{abstract}

\section{Introduction}
The need to adopt post-quantum algorithms for secure systems design is required due to continued advances in the development of quantum computing technology.  The US National Institute of Standards and Technology (NIST) recommends that deployers of systems that will be in operation for 10 years or more should adopt post-quantum methods \emph{now}.  NIST specifically recommends the use of Stateful Hash-Based Signatures (SHBS) as a leading candidate for a range of post-quantum applications including software updates.\cite{800-208}  This recommendation comes with specific operational requirements that are needed to address the unique challenges when adopting SHBS algorithms.\cite{NSA-CNSA-2022r2}  The two SHBS algorithms that have been approved for use by NIST offer a large number of configuration parameter sets that have a significant impact on the performance and storage requirements introduced when switching from a conventional digital signature algorithm to a SHBS method.

The intended use of this benchmark data is to provide a reference that can be used to compare relative performance when choosing among the various SHBS parameter options.  Data is provided for XMSS and LMS, the two approved SHBS algorithms.  Comparable metrics are also provided for SPHINCS+, a non-stateful algorithm and ECDSA on the P256 curve.  These non-stateful alternatives are included for comparison.

Note that this data was generated with a specific interest in understanding the impact of SHBS algorithm parameters on digital signature distribution and validation.  The intended use case for this data is to consider a collection of distributed systems which must accept and validate signatures that are generated at a central location.  In this scenario, the burden of keypair generation, state management, and signature generation are all centralized and can take advantage of specialized hardware.  The challenge is in adapting systems in the field to accept and validate the resulting signatures.  

Table \ref{tab:compare} shows a high-level comparison of ECDSA along with the maximum and minimum values for XMSS and LMS across all approved parameter values.  ECDSA on P256 was selected as a reference since it is frequently used in many current deployments on embedded and IoT systems.  Detailed results for XMSS and XMSS-MT are described in Section \ref{appendix_lms}.  Comparable results for LMS and HSS are described in Section \ref{appendix_lms}.  Results for ECDSA and SPHINCS+ are in Sections \ref{ECDSA Benchmark Data} \ref{appendix_sphincs} respectively.

The data provided is intended for use in making relative comparisons between different options.  All of the validation timing data was collected on the same platform with the same compiler options.  Performance of these algorithms may be very different on systems with a different processor architecture or built with different compiler options.  However, the relative performance of one parameter set when compared to the other options is expected to remain relatively consistent across systems.  

\begin{table}[tbp] \centering
  \begin{tabular}{lccccc}
    & \textbf{\begin{tabular}[c]{@{}c@{}}ECDSA\\ P-256\end{tabular}} &
    \multicolumn{2}{c}{
      \textbf{\begin{tabular}[c]{@{}c@{}}XMSS\end{tabular}}} &
    \multicolumn{2}{c}{
      \textbf{\begin{tabular}[c]{@{}c@{}}LMS\end{tabular}}} \\
    &     & Min   & Max    & Min     & Max    \\
    Number of Signatures & NA & $2^{10}$ & $2^{60}$ & $2^{5}$ & $2^{40}$\\
    Signature Size (KB) & 0.06 & 1.46 & 27.04 & 1.27 & 9.11\\
    Public Key Size (B) & 65 & 68 & 132 & 60 & 60\\
    Validation Time (M-Cycles) & 0.23 & 2.14 & 54.67 & 0.35 & 28.17
\end{tabular}
  \caption{Comparison of ECDSA with XMSS and LMS}
  \label{tab:compare}
\end{table}

On a general purpose computing platform the differences among the various parameter sets may have a negligible impact on the overall system performance and security.  On a system with a fast network, vast storage, and significantly underutilized computational capacity coupled with relatively infrequent software updates, even a highly inefficient set of parameters will have no noticeable impact on the overall system.  The same is not true for highly optimized embedded systems.  In some cases, these systems are designed to operate on the edge of a key parameter such as battery life or system storage.  Some systems operate in extreme environments where the additional computational, communication, or storage requirements imposed by some parameter options would cause a previously viable system to fall short of the design goals.  It is therefore valuable to consider the trade space introduced by the SHBS parameter sets.  

\subsection{Number of Signatures}
A core property of any SHBS algorithm is that a signing key can only be used a limited number of times.  This means that each key can only sign a limited number of messages.  This has two impacts.  First, the system state must be maintained to keep track of which components of the private key have been used and which ones are still available.  Second, the overall system must account for the fact that each signing key is a limited resource that must be replaced when it reaches the maximum number of signatures allowed for the selected SHBS parameters.  

\subsection{Public Key Size}
Public key size impacts the storage required and may impact communication bandwidth in an event where a new public key must be distributed.  In the most basic software update system, a single public key is installed during manufacturing and it is used to validate updates over the lifetime of the device, so the size of the public key only impacts the internal storage required.  In a system where new public keys can be distributed to remote systems, the size of each key will impact the network bandwidth needed to distribute new keys.  

\subsection{Signature Size}
Every secure message must be accompanied by at least one digital signature to validate the integrity and authenticity of the message content.  Large digital signatures require additional network resources to distribute and store signed content.

\subsection{Computational Cost of Validation}
Many embedded systems have a limited budget of compute resources that can be allocated to the software update process.  The computational effort needed to validate software update signatures may interfere with critical real-time control system functions, or require that they be scheduled during limited periods allocated for maintenance functions.  In the case of a system that is battery operated, computation required to validate an update may have a direct impact on the battery life of the system.

\section{XMSS and XMSS\textsuperscript{MT} Benchmark Data}
\label{appendix_xmss}
Table \ref{tab:XMSS_behcnmarks} contains benchmark data for the XMSS and XMSS\textsuperscript{MT} signature methods.  This data was generated using the \texttt{libOQS} library version \texttt{0.11.1-dev} \cite{StebilaMosca2017}.  The ``Algorithm ID'' matches the algorithm identifiers defined in the \texttt{libOQS} code.  Values for signature, public, and secret key lengths are reported by the library.

Validation time was calculated by executing the validation routine on an x86 system.  The values shown are an approximation of the total cycles required for a single message valuation, shown as millions of cycles.  The estimate was calculated by averaging the in-process time for 1,000 validation operations and converting the time to cycles using the system reported CPU frequency.  

\newgeometry{left=1cm,right=1cm}
\begin{table}[tbp] \centering
    \begin{tabular}{|l|r|r|r|r|r|}
    \multicolumn{1}{c}{\textbf{Algorithm ID}} &
    \multicolumn{1}{c}{\textbf{
        \begin{tabular}[c]{@{}c@{}}Number of \\Signatures \\$log_2(N)$
        \end{tabular}}} &
    \multicolumn{1}{c}{\textbf{
        \begin{tabular}[c]{@{}c@{}}Signature\\Length\\(KB)
        \end{tabular}}} &
    \multicolumn{1}{c}{\textbf{
        \begin{tabular}[c]{@{}c@{}}Public\\Key\\Length\\(B)
        \end{tabular}}} &
    \multicolumn{1}{c}{\textbf{
        \begin{tabular}[c]{@{}c@{}}Secret\\Key\\Length\\(B)
        \end{tabular}}} &
    \multicolumn{1}{c}{\textbf{
        \begin{tabular}[c]{@{}c@{}}Validation\\Time\\(M-Cycles)
        \end{tabular}}} \\ \hline
    XMSS-SHA2\_10\_256 & 10 & 2.44 & 68 & 1,377 & 4.62 \\ \hline
    XMSS-SHA2\_16\_256 & 16 & 2.63 & 68 & 2,097 & 4.28 \\ \hline
    XMSS-SHA2\_20\_256 & 20 & 2.75 & 68 & 2,577 & 4.63 \\ \hline
    XMSS-SHAKE\_10\_256 & 10 & 2.44 & 68 & 1,377 & 2.68 \\ \hline
    XMSS-SHAKE\_16\_256 & 16 & 2.63 & 68 & 2,097 & 2.88 \\ \hline
    XMSS-SHAKE\_20\_256 & 20 & 2.75 & 68 & 2,577 & 2.80 \\ \hline
    XMSS-SHA2\_10\_512 & 10 & 8.88 & 132 & 2,657 & 11.08 \\ \hline
    XMSS-SHA2\_16\_512 & 16 & 9.25 & 132 & 4,049 & 11.75 \\ \hline
    XMSS-SHA2\_20\_512 & 20 & 9.50 & 132 & 4,977 & 11.47 \\ \hline
    XMSS-SHAKE\_10\_512 & 10 & 8.88 & 132 & 2,657 & 10.52 \\ \hline
    XMSS-SHAKE\_16\_512 & 16 & 9.25 & 132 & 4,049 & 11.58 \\ \hline
    XMSS-SHAKE\_20\_512 & 20 & 9.50 & 132 & 4,977 & 11.81 \\ \hline
    XMSS-SHA2\_10\_192 & 10 & 1.46 & 52 & 1,057 & 3.08 \\ \hline
    XMSS-SHA2\_16\_192 & 16 & 1.60 & 52 & 1,609 & 3.03 \\ \hline
    XMSS-SHA2\_20\_192 & 20 & 1.69 & 52 & 1,977 & 3.00 \\ \hline
    XMSS-SHAKE256\_10\_192 & 10 & 1.46 & 52 & 1,057 & 2.14 \\ \hline
    XMSS-SHAKE256\_16\_192 & 16 & 1.60 & 52 & 1,609 & 2.37 \\ \hline
    XMSS-SHAKE256\_20\_192 & 20 & 1.69 & 52 & 1,977 & 2.36 \\ \hline
    XMSS-SHAKE256\_10\_256 & 10 & 2.44 & 68 & 1,377 & 3.21 \\ \hline
    XMSS-SHAKE256\_16\_256 & 16 & 2.63 & 68 & 2,097 & 3.11 \\ \hline
    XMSS-SHAKE256\_20\_256 & 20 & 2.75 & 68 & 2,577 & 2.90 \\ \hline
    XMSSMT-SHA2\_20/2\_256 & 20 & 4.85 & 68 & 6,002 & 9.26 \\ \hline
    XMSSMT-SHA2\_20/4\_256 & 20 & 9.03 & 68 & 10,942 & 17.72 \\ \hline
    XMSSMT-SHA2\_40/2\_256 & 40 & 5.47 & 68 & 9,604 & 8.94 \\ \hline
    XMSSMT-SHA2\_40/4\_256 & 40 & 9.66 & 68 & 15,256 & 19.45 \\ \hline
    XMSSMT-SHA2\_40/8\_256 & 40 & 18.04 & 68 & 24,520 & 36.77 \\ \hline
    XMSSMT-SHA2\_60/3\_256 & 60 & 8.20 & 68 & 16,633 & 13.49 \\ \hline
    XMSSMT-SHA2\_60/6\_256 & 60 & 14.48 & 68 & 24,511 & 26.69 \\ \hline
    XMSSMT-SHA2\_60/12\_256 & 60 & 27.04 & 68 & 38,099 & 54.67 \\ \hline
    XMSSMT-SHAKE\_20/2\_256 & 20 & 4.85 & 68 & 6,002 & 6.81 \\ \hline
    XMSSMT-SHAKE\_20/4\_256 & 20 & 9.03 & 68 & 10,942 & 11.30 \\ \hline
    XMSSMT-SHAKE\_40/2\_256 & 40 & 5.47 & 68 & 9,604 & 5.77 \\ \hline
    XMSSMT-SHAKE\_40/4\_256 & 40 & 9.66 & 68 & 15,256 & 12.08 \\ \hline
    XMSSMT-SHAKE\_40/8\_256 & 40 & 18.04 & 68 & 24,520 & 24.06 \\ \hline
    XMSSMT-SHAKE\_60/3\_256 & 60 & 8.20 & 68 & 16,633 & 8.89 \\ \hline
    XMSSMT-SHAKE\_60/6\_256 & 60 & 14.48 & 68 & 24,511 & 17.98 \\ \hline
    XMSSMT-SHAKE\_60/12\_256 & 60 & 27.04 & 68 & 38,099 & 37.33 \\ \hline
    \end{tabular}
    \caption{XMSS and XMSSMT benchmark data}
    \label{tab:XMSS_behcnmarks}
\end{table}
\restoregeometry

\section{LMS and HSS Benchmark Data}
\label{appendix_lms}
Table \ref{tab:LMS_behcnmarks} contains benchmark data for the LMS and HSS signature methods.  This data was generated using the \texttt{libOQS} library, version \texttt{0.11.1-dev}\cite{StebilaMosca2017}.  

Validation time was calculated through executing the validation routine on an x86 system.  The values shown are an approximation of the total cycles required for a single message validation, shown as millions of cycles.  The estimate was calculated by averaging the in-process time for 1,000 validation operations and converting the time to cycles using the system reported CPU frequency.  

Note that data is missing for 3 of the parameter options, specifically LMS\_SHA256\_H25\_W1, LMS\_SHA256\_H25\_W2, and LMS\_SHA256\_H25\_W8.  This is due to issues encountered with the library in executing these options.  

\newgeometry{left=1cm,right=1cm}
\begin{table}[tbp] \centering
    \begin{tabular}{|l|r|r|r|r|r|}
    \multicolumn{1}{c}{\textbf{Algorithm ID}} &
    \multicolumn{1}{c}{\textbf{
        \begin{tabular}[c]{@{}c@{}}Number of \\Signatures \\$log_2(N)$
        \end{tabular}}} &
    \multicolumn{1}{c}{\textbf{
        \begin{tabular}[c]{@{}c@{}}Signature\\Length\\(KB)
        \end{tabular}}} &
    \multicolumn{1}{c}{\textbf{
        \begin{tabular}[c]{@{}c@{}}Public\\Key\\Length\\(B)
        \end{tabular}}} &
    \multicolumn{1}{c}{\textbf{
        \begin{tabular}[c]{@{}c@{}}Secret\\Key\\Length\\(B)
        \end{tabular}}} &
    \multicolumn{1}{c}{\textbf{
        \begin{tabular}[c]{@{}c@{}}Validation\\Time\\(M-Cycles)
    \end{tabular}}} \\ \hline
        LMS\_SHA256\_H5\_W1 & 5 & 8.48 & 60 & 64 & 0.35 \\ \hline
        LMS\_SHA256\_H5\_W2 & 5 & 4.36 & 60 & 64 & 0.36 \\ \hline
        LMS\_SHA256\_H5\_W4 & 5 & 2.30 & 60 & 64 & 0.71 \\ \hline
        LMS\_SHA256\_H5\_W8 & 5 & 1.27 & 60 & 64 & 5.60 \\ \hline
        LMS\_SHA256\_H10\_W1 & 10 & 8.64 & 60 & 64 & 0.38 \\ \hline
        LMS\_SHA256\_H10\_W2 & 10 & 4.52 & 60 & 64 & 0.41 \\ \hline
        LMS\_SHA256\_H10\_W4 & 10 & 2.45 & 60 & 64 & 0.70 \\ \hline
        LMS\_SHA256\_H10\_W8 & 10 & 1.42 & 60 & 64 & 6.43 \\ \hline
        LMS\_SHA256\_H15\_W1 & 15 & 8.80 & 60 & 64 & 0.39 \\ \hline
        LMS\_SHA256\_H15\_W2 & 15 & 4.67 & 60 & 64 & 0.40 \\ \hline
        LMS\_SHA256\_H15\_W4 & 15 & 2.61 & 60 & 64 & 0.73 \\ \hline
        LMS\_SHA256\_H15\_W8 & 15 & 1.58 & 60 & 64 & 5.78 \\ \hline
        LMS\_SHA256\_H20\_W1 & 20 & 8.95 & 60 & 64 & 0.40 \\ \hline
        LMS\_SHA256\_H20\_W2 & 20 & 4.83 & 60 & 64 & 0.38 \\ \hline
        LMS\_SHA256\_H20\_W4 & 20 & 2.77 & 60 & 64 & 0.71 \\ \hline
        LMS\_SHA256\_H20\_W8 & 20 & 1.73 & 60 & 64 & 6.82 \\ \hline
        LMS\_SHA256\_H25\_W1 & 25 & 9.11 & 60 & 64 & ~ \\ \hline
        LMS\_SHA256\_H25\_W2 & 25 & 4.98 & 60 & 64 & ~ \\ \hline
        LMS\_SHA256\_H25\_W4 & 25 & 2.92 & 60 & 64 & 0.75 \\ \hline
        LMS\_SHA256\_H25\_W8 & 25 & 1.89 & 60 & 64 & ~ \\ \hline
        LMS\_SHA256\_H5\_W8\_H5\_W8 & 10 & 2.58 & 60 & 64 & 12.77 \\ \hline
        LMS\_SHA256\_H10\_W4\_H5\_W8 & 15 & 3.77 & 60 & 64 & 16.02 \\ \hline
        LMS\_SHA256\_H10\_W8\_H5\_W8 & 15 & 2.74 & 60 & 64 & 28.17 \\ \hline
        LMS\_SHA256\_H10\_W2\_H10\_W2 & 20 & 9.08 & 60 & 64 & 1.57 \\ \hline
        LMS\_SHA256\_H10\_W4\_H10\_W4 & 20 & 4.96 & 60 & 64 & 2.86 \\ \hline
        LMS\_SHA256\_H10\_W8\_H10\_W8 & 20 & 2.89 & 60 & 64 & 24.51 \\ \hline
        LMS\_SHA256\_H15\_W8\_H5\_W8 & 20 & 2.89 & 60 & 64 & 24.18 \\ \hline
        LMS\_SHA256\_H15\_W8\_H10\_W8 & 25 & 3.05 & 60 & 64 & 15.37 \\ \hline
        LMS\_SHA256\_H15\_W8\_H15\_W8 & 30 & 3.21 & 60 & 64 & 17.20 \\ \hline
        LMS\_SHA256\_H20\_W8\_H5\_W8 & 25 & 3.05 & 60 & 64 & 15.81 \\ \hline
        LMS\_SHA256\_H20\_W8\_H10\_W8 & 30 & 3.21 & 60 & 64 & 19.08 \\ \hline
        LMS\_SHA256\_H20\_W8\_H15\_W8 & 35 & 3.36 & 60 & 64 & 24.19 \\ \hline
        LMS\_SHA256\_H20\_W8\_H20\_W8 & 40 & 3.52 & 60 & 64 & 11.53 \\ \hline
    \end{tabular}
    \caption{LMS and HSS benchmark data}
    \label{tab:LMS_behcnmarks}
\end{table}
\restoregeometry

\section{ECDSA Benchmark Data}
\label{appendix_ecdsa}
Table \ref{tab:ECDSA_behcnmarks} contains benchmark data for the ECDSA signature method.  This data was generated using \texttt{openssl} version \texttt{3.0.13} on the same x86 platform as the other benchamrks.  

\begin{table}[tbp] \centering
    \begin{tabular}{|l|r|r|r|}
    \multicolumn{1}{c}{\textbf{Algorithm ID}} &
    \multicolumn{1}{c}{\textbf{
        \begin{tabular}[c]{@{}c@{}}Signature\\Length\\(KB)
        \end{tabular}}} &
    \multicolumn{1}{c}{\textbf{
        \begin{tabular}[c]{@{}c@{}}Public\\Key\\Length\\(B)
        \end{tabular}}} &
    \multicolumn{1}{c}{\textbf{
        \begin{tabular}[c]{@{}c@{}}Validation\\Time\\(M-Cycles)
    \end{tabular}}} \\ \hline
    ECDSA-P256 & 0.06 & 32 & 0.23 \\ \hline
    \end{tabular}
    \caption{ECDSA benchmark data}
    \label{tab:ECDSA_behcnmarks}
\end{table}

\section{SPHINCS+ Benchmark Data}
\label{appendix_sphincs}
Table \ref{tab:SPHINCS_behcnmarks} contains benchmark data for the SPHINCS+ signature method.  This data was generated using the \texttt{libOQS} library \cite{StebilaMosca2017}.  The ``Algorithm ID'' matches the algorithm identifiers defined in the \texttt{libOQS} code, version \texttt{0.11.1-dev}.  Values for signature, public, and secret key lengths are reported by the library.

Validation time was calculated by executing the validation routine on an x86 system.  The values shown are an approximation of the total cycles required for a single message valuation, shown as millions of cycles.  The estimate was calculated by averaging the in-process time for 1,000 validation operations and converting the time to cycles using the system reported CPU frequency.  

\begin{table}[tbp] \centering
    \begin{tabular}{|l|r|r|r|}
    \multicolumn{1}{c}{\textbf{Algorithm ID}} &
    \multicolumn{1}{c}{\textbf{
        \begin{tabular}[c]{@{}c@{}}Signature\\Length\\(KB)
        \end{tabular}}} &
    \multicolumn{1}{c}{\textbf{
        \begin{tabular}[c]{@{}c@{}}Public\\Key\\Length\\(B)
        \end{tabular}}} &
    \multicolumn{1}{c}{\textbf{
        \begin{tabular}[c]{@{}c@{}}Validation\\Time\\(M-Cycles)
    \end{tabular}}} \\ \hline
    SPHINCS+-SHA2-128f-simple & 16.68 & 32 & 1.904  \\ \hline
    SPHINCS+-SHA2-128s-simple &  7.68 & 32 & 0.844  \\ \hline
    SPHINCS+-SHA2-192f-simple & 34.83 & 48 & 2.839  \\ \hline
    SPHINCS+-SHA2-192s-simple & 15.84 & 48 & 1.190  \\ \hline
    SPHINCS+-SHA2-256f-simple & 48.69 & 64 & 2.947  \\ \hline
    SPHINCS+-SHA2-256s-simple & 29.09 & 64 & 1.637  \\ \hline
    SPHINCS+-SHAKE-128f-simple & 16.69 & 32 & 3.081  \\ \hline
    SPHINCS+-SHAKE-128s-simple &  7.67 & 32 & 1.223  \\ \hline
    SPHINCS+-SHAKE-192f-simple & 34.83 & 48 & 4.719  \\ \hline
    SPHINCS+-SHAKE-192s-simple & 15.84 & 48 & 1.705  \\ \hline
    SPHINCS+-SHAKE-256f-simple & 48.69 & 64 & 4.361  \\ \hline
    SPHINCS+-SHAKE-256s-simple & 29.09 & 64 & 2.365  \\ \hline
    \end{tabular}
    \caption{ECDSA benchmark data}
    \label{tab:SPHINCS_behcnmarks}
\end{table}

\bibliographystyle{plain} 
\bibliography{Romansky-SHBS_benchmark} 

\end{document}